\documentclass[10pt]{article}%
\usepackage{amsfonts}
\usepackage{amsmath}
\usepackage{amssymb}
\usepackage{graphicx}
\usepackage[usenames]{color}%
\def \be {\begin{equation}}
\def \ee {\end{equation}}
\def \bea {\begin{eqnarray}}
\def \eea {\end{eqnarray}}
\def \nn {\nonumber}

\def \rr {\raise.35ex\hbox{\small $\prime$}\kern-.17em{\mbox{\large $\imath$}}}
\def \del {\partial}
\def \dels {\partial\kern-.5em / \kern.5em}
\def \As {{A\kern-.5em / \kern.5em}}
\def \Ds {D\kern-.7em / \kern.5em}

\def \a {\alpha}
\def \ap {\alpha}

\def \II {I\hspace{-.15em}I\hspace{.1em}}
\def \III {I\hspace{-.15em}I\hspace{-.15em}I\hspace{.1em}}

\def \Z {{\mathbb{Z}}}

\begin{document}
\begin{titlepage}
\begin{center}
\hfill hep-th/0509009\\
\vskip .5in
\textbf{\large
Comments on the high energy limit of
bosonic open string theory
}
\vskip .5in
Chuan-Tsung Chan $^1$, Pei-Ming Ho $^2$, Jen-Chi~Lee $^3$, \\
Shunsuke Teraguchi $^4$, Yi Yang $^3$
\vskip 10pt
{\small $^1$ Physics Division, National Center for Theoretical Sciences,
Hsinchu, Taiwan, R.O.C.} \\
{\small $^2$ Department of Physics, National Taiwan University, Taipei,
Taiwan, R.O.C.} \\
{\small $^3$ Department of Electrophysics, National Chiao-Tung University,
Hsinchu, Taiwan, R.O.C.} \\
{\small $^4$ Physics Division, National Center for Theoretical Sciences,
Taipei, Taiwan, R.O.C.}
\vskip .2in
\sffamily{
ctchan@phys.cts.nthu.edu.tw\\
pmho@ntu.edu.tw\\
jcclee@cc.nctu.edu.tw\\
teraguch@phys.ntu.edu.tw\\
yiyang@mail.nctu.edu.tw
}
\vspace{60pt}
\end{center}
\begin{abstract}
In previous works, ratios among four-point scattering amplitudes at the leading order
in the high-energy limit were derived for the bosonic open string theory.
The derivation was based on Ward identities derived from the decoupling of zero-norm states and was purely algebraic.
The only assumption of the derivation was that the momentum polarization
can be approximated by the longitudinal polarization at high energies.
In this paper, using the decoupling of spurious states,
we reduce this assumption to a much weaker one
which can be easily verified by simple power counting in most cases.
For the special cases which are less obvious,
we verify the new assumption for an example by saddle-point approximation.
We also provide a new perspective to our previous results
in terms of DDF states. In particular, we show that, by using DDF states, one can easily see that there
is only one independent high energy scattering amplitude for each fixed mass level.
\end{abstract}
\end{titlepage}
\setcounter{footnote}{0}

\section{Introduction}

One of the prominent characteristics of string theory is its huge spacetime
gauge symmetry. In the usual world-sheet approach, this gauge symmetry is
represented by zero-norm states in the spectrum and string interactions
preserve this structure. It is reasonable to believe that, at least
perturbatively, this huge gauge symmetry together with Lorentz symmetry govern
the theory in flat spacetime.

In \cite{Gross}, it was conjectured that string theory also possesses a huge
hidden symmetry which relates all string oscillation modes to one another,
such that the S-matrix is determined by the scattering amplitudes among, say,
tachyons for bosonic strings or dilatons for superstrings. Unfortunately, not
much of this hidden symmetry has been understood.

In general, it is plausible that a hidden symmetry becomes manifest in the
high-energy limit \cite{GM},
where massive particles become effectively massless. The
observation that only the above-mentioned gauge symmetry governs the theory
suggests that the hidden symmetry is in some sense hidden in the gauge
symmetry\footnote{A possible connection between these symmetries is that the
hidden symmetry is needed for the self-consistency of gauge-invariant
interactions, as it is highly nontrivial for higher spin gauge theories to
have self-consistent interactions.}. If this is the case, taking the
high-energy limit of the gauge symmetry might be a proper approach in order to
obtain useful information about the conjectured hidden symmetry. Our previous
works \cite{ChanLee1,CHL,CHLTY} were such efforts.
For other attempts, see \cite{WS,others}.

In the papers \cite{ChanLee1,CHL}, we derived some linear relations
among high-energy 4-point scattering amplitudes up to the third mass level by
considering the high-energy limit of zero-norm states. Furthermore, in the paper
\cite{CHLTY}, we generalized previous results to all mass levels. The strategy
of our previous approach is the following. First, using the decoupling of
zero-norm states from amplitudes, one obtains some linear relations (stringy
Ward identities) among the amplitudes for unphysical states\footnote{Of
course, ``amplitude'' for unphysical states is not well-defined. They depend
on the gauge-fixing prescription for world sheet symmetry.}. Note that
zero-norm states themselves do not relate physically inequivalent particle
states, therefore we can not obtain any physically meaningful relations at
this point. However, after taking the high-energy limit of these relations,
something special happens. Recall that 1-string states are characterized by a
choice of polarizations in 26 dimensions. In this limit,
momentum
polarization and longitudinal polarization approach each other, and the
transverse directions can be neglected except for the direction parallel to
the scattering plane under consideration. Therefore, if we ignore the
$1/E^{2}$ effects, the system looks effectively two-dimensional\footnote{This is for the
case of 4-point function.} and the above stringy Ward identities can relate
amplitudes which now only involve two polarizations. Using these linear
relations in the high-energy limit, one can obtain physically meaningful
results. Remarkably, all the high-energy leading amplitudes for relevant
physical states are completely solved and can be related to that of four tachyons.

However, there is a loophole in the above argument.
While we only focused on the leading behavior of amplitudes, sometimes an
amplitude vanishes accidentally at the presumed leading order, and the true
leading order is lower than the naive expectation. In this case, we can not
fully justify the omission of the $1/E^{2}$ effects (which are actually at the
true leading order), and the replacement of the momentum polarization by the
longitudinal polarization may not be a good approximation. As a simple
example, using the notation in Eq.(\ref{ePLT}), the difference between $e^{P}
\cdot\mathcal{T}$ and $e^{L} \cdot\mathcal{T}$ is at the leading order if
$\mathcal{T} \propto e^{L}$ or $e^{P}$ although the difference between $e^{P}$
and $e^{L}$ is subleading in the high energy limit. Therefore, strictly
speaking, our previous argument needs further justification. Roughly
speaking, it was based on an assumption on the smoothness of the high-energy
limit, as we explained in the paper \cite{CHL}. Although this assumption is
highly nontrivial from the viewpoint of ordinary field theories, the final
results have been verified independently by direct computations of four-point
functions. The main aim of this paper is to fill in this possible loophole.
Instead of starting with stringy Ward identities which are derived from the
decoupling of zero-norm states, we utilize the decoupling of spurious states
here. With a much weaker assumption, we can justify the omission of the
$1/E^{2}$ effects on the way to showing the irrelevance of other states, that
is, the momentum polarization can be replaced by the longitudinal polarization
at high energies. Besides this, the derivations of linear relations are quite
similar to those in \cite{CHLTY} based on the decoupling of high-energy
zero-norm states. Though the new assumption for the proof of our results might
seem always valid at first sight, if we consider multi-tensor scattering
amplitudes, it is not always trivial, at least not until explicitly checked.
For two-tensor cases, we shall show one example, and check that the assumption
is valid. Based on this example, we argue that this assumption is valid for
generic cases.

{}To summarize, we can now show with better rigor that
the high-energy amplitudes of bosonic open strings are linearly related.
Furthermore, only one independent function for high-energy amplitudes exists
at each fixed mass level. This result suggests that effectively only one
physical state survives in this high-energy limit
at every mass level. 
In principle, using our results,
one can always perform a change of the basis for physical states
so that the ratios among the high-energy amplitudes become $1:0:\cdots:0$.
Moreover, by properly choosing the gauge, we can explicitly see this remarkable fact.
We shall also comment on this issue in this paper.

\section{Rederivation of the general formula and justification of the
replacement of $P$ by $L$}

\label{Spurious}

\subsection{The high-energy limit}

Before detailing the proof of our result, we briefly summarize our procedure
of taking the high-energy limit. See \cite{CHLTY} for more details. We only
consider 4-point scattering amplitudes of open strings in bosonic string
theory for simplicity. These 4-point amplitudes depend on the center-of-mass
energy $E_{\mathrm{cm}}$, the scattering angle $\phi$ and the choices of four
oscillating modes of strings. We take the limit\footnote{Namely, our arguments
are valid only in the region where $E_{\mathrm{cm}}\gg2(n-1)$. We are using
the convention of $\ap^{\prime}=1/2$.} of $E_{\mathrm{cm}}$ going to infinity
with the scattering angle $\phi$ fixed. Three of the particles are fixed and
the mass level of the last particle is a fixed integer $n$. We will use the
terminology ``family'' to represents a class of particles which are at the
same mass level $n$. Our question is how the leading behavior of the
amplitudes will change when we replace one of the particles by another
particle in the same family. Now we will focus on the string state of the last
particle. To specify the polarizations of this state, we use the following
basis for 26 dimensions
\begin{equation}
e^{P} = \frac{1}{m}(\sqrt{p^{2}+m^{2}}, p, 0, \cdots, 0), \hspace{0.3cm} e^{L}
= \frac{1}{m}(p,\sqrt{p^{2}+m^{2}}, 0, \cdots, 0), \hspace{0.3cm} e^{T_{i}} =
(0,0,\cdots,1, \cdots), \label{ePLT}%
\end{equation}
where $m$ is the mass of the state at issue given by $\sqrt{2(n-1)}$. It will
be convenient to introduce two more vectors
\begin{equation}
e^{T}=(0,0,\cdots,0,1), \hspace{0.3cm} e^{(L-P)}=\frac{p-\sqrt{p^{2}+m^{2}}%
}{m}(1,1, 0, \cdots, 0)\simeq-\frac{m}{2p}(1,1,0,\cdots,0),
\end{equation}
where $e^{T}$ is the transverse vector which is parallel to the scattering
plane\footnote{In this paper, we set the scattering plane on the $X^{0}%
-X^{25}$ plane.}. {}From now on, we reserve the notation $e^{T_{i}}$ for the
transverse polarizations which are perpendicular to the scattering plane.
$e^{(L-P)}$ is a null vector which is defined by the difference between
$e^{P}$ and $e^{L}$. The inner product of a vector $V^{\mu}$ with these unit
vectors $e^{A}_{\mu}$ ($A=P, L, T, T_{i}$ or $(L-P)$) will be denoted by $V^{A}$.

\subsection{Definitions and assumption}

In the rest of this section, we give a better proof of our main result in
\cite{CHLTY}. This will be achieved in several steps. First, we will assign
naive energy dimensions to each oscillation modes based on concrete
calculation of scattering amplitudes. Second, we will have to make a minor
assumption about how 4-point functions really scale with energy $E$ in the
high-energy limit, namely, we assume that, for some class of states, we can
trust the power counting of the naive energy dimensions. This assumption can
sometimes be easily checked using the saddle-point approximation. {}From the
decoupling of spurious states and this assumption, we can tell which states
should not contribute to scattering amplitudes in the high-energy limit. This
information will allow us to freely replace the polarization $P$ by $L$ in the
linear relations of 4-point functions obtained from the decoupling of spurious
states. The last step is to show that these linear relations lead to our
earlier result \cite{CHLTY}, which is the explicit expression of ratios among
scattering amplitudes.

As we just mentioned, we use the decoupling of spurious states from physical
states to rederive our previous result in a rigorous manner. We only need two
Virasoro operators
\begin{align}
L_{-1}  &  = \frac{1}{2} \sum_{n\in\Z} \a_{-1+n}\cdot\a_{-n} = \hat
m\ap_{-1}^{P}+\ap_{-2}\cdot\ap_{1}+\cdots,\\
L_{-2}  &  = \frac{1}{2} \sum_{n\in\Z} \a_{-2+n}\cdot\a_{-n} = \frac{1}{2}
\ap_{-1}\cdot\ap_{-1}+\hat m\ap_{-2}^{P}+\ap_{-3}\cdot\ap_{1}+\cdots,
\end{align}
to generate all spurious states. Each oscillator $\ap_{-m}^{A}$ corresponds to
a factor of $\frac{1}{(m-1)!}\del^{m} X^{A}$ in the vertex operator. The
operator $\del^{m} X^{\mu}$ can contract with the exponent $i k\cdot X$ of
another vertex operator in the correlation function to produce a factor of
$k^{\mu}$. This is the leading order contribution of the factor $\del^{m}
X^{\mu}$ to the correlation function, and it scales like $E^{1}$ at high
energies. So we assign a dimension $1$ to $\del^{m} X^{\mu}$. Similarly we
assign dimensions to the polarization vectors. Combining the dimensions of
$X^{\mu}$ and $e^{A}_{\mu}$, we associate a naive dimension to every
oscillator
\begin{align}
\ap_{-m}^{P} \rightarrow2, \quad\ap_{-m}^{L} \rightarrow2, \quad\ap_{-m}^{T}
\rightarrow1, \quad\ap_{-m}^{T_{i}} \rightarrow0, \quad\ap_{-m}^{(L-P)}
\rightarrow0. \label{naivedim}%
\end{align}
However, terms at the naive leading order may happen to cancel (this happens
whenever $\del X^{P}$ or $\del X^{L}$ is involved
\cite{ChanLee1,CHL,CHLTY}) and the true leading order may be lower
(but never higher).

Another notion that will be helpful is the naive dimension of a state, which
is the sum of the naive dimensions of all creation operators needed to create
the state from vacuum. We symbolically represent a generic state at level $n$
(with mass $\hat{m}= \sqrt{2(n-1)}$) and naive dimension $d$ as $| n,d
\rangle$. As an example of using this notation, we have
\begin{equation}
L_{-1}| n-1,d \rangle=\hat m \ap_{-1}^{P}| n-1,d \rangle+| n,d \rangle.
\end{equation}
Let us now state our assumption.

\noindent\emph{Assumption}

In the following we will assume that the true energy order of the amplitude
for the state
\begin{equation}
|n\rangle= \overbrace{\ap_{-1}^{T}\cdots\ap_{-1}^{T}}^{n}| 0 \rangle,
\label{assumption}%
\end{equation}
(with other particles fixed) is greater than the amplitude for any state $|
n,d \rangle$ whose naive dimension $d$ is less than $n$.

Note that this is the only input of our derivation apart from the decoupling
of spurious states. Using the saddle-point method, one can immediately
conclude that the assumption is correct for 4-point functions when the
prefactors of the other 3 vertices do not contain $\del X^{L}$. Therefore the
assumption only needs to be checked when $\del X^{L}$ does appear in one or
more of the other 3 vertices. We do not have a rigorous proof, but the
assumption holds for all examples we have checked. (See sec.
\ref{Check of the assumption}.)

In fact, our algebraic proof can be applied to generic $N$-point functions.
For $N$-point functions, we fix $(N-1)$ vertices such that the amplitude is
not suppressed when the varying vertex is chosen to be $|n\rangle$. The same
procedure given below will allow us to find the ratios between the $N$-point
function for $|n\rangle$ and certain other choices of states at the same
level. However, unique solution for all possible choices of vertices can be
derived only for 4-point functions because in other cases we can not ignore
all other transverse polarizations.

\subsection{Proof of irrelevance}

The first step is to find those states which are subleading compared to
$|n\rangle$, so that we can ignore them later. We first show that a state is
subleading if the total number of $\a^{P}_{-m}$ and $\a^{L}_{-m}$ is odd.

\subsubsection{Irrelevance of the states with only one $\ap_{-1}^{P}$}

{}To begin, we prove that states involving a single factor of $\ap_{-1}^{P}$ are
subleading. Consider a class of spurious states generated by $L_{-1}$
\begin{align}
L_{-1}| n-1,n-1 \rangle=\hat m \ap_{-1}^{P}| n-1,n-1 \rangle+\underbrace{|
n,n-1 \rangle}_{\mathrm{irrelevant } }.
\end{align}
Due to the assumption (\ref{assumption}) above, the state $| n,n-1 \rangle$ is
not at the leading order and can be ignored. The decoupling of the spurious
state implies that
\begin{align}
\ap_{-1}^{P}| n-1,n-1 \rangle\rightarrow\mathrm{irrelevant } , \label{d1p1}%
\end{align}
though the naive dimension of this state is $n+1$.

\subsubsection{Irrelevance of the states with three $\ap_{-1}^{P}$}

The next set of spurious states we consider is
\begin{align}
L_{-1}\ap_{-1}^{P}| n-2,n-3 \rangle=\hat m \ap_{-1}^{P} \ap_{-1}^{P}| n-2,n-3
\rangle+\underbrace{\ap_{-2}^{P}| n-2,n-3 \rangle}_{\mathrm{irrelevant }
}+\underbrace{\ap_{-1}^{P}| n-1,n-3 \rangle}_{\mathrm{irrelevant } }.
\label{d2p1}%
\end{align}
The last two terms are of naive leading order $(n-1)$, implying that the first
term on the right hand side is decoupled in the high energy limit, despite the
fact that it has a naive dimension of $(n+1)$. Similarly, we have
\begin{align}
L_{-1}\ap_{-1}^{P}\ap_{-1}^{P}| n-3,n-3 \rangle=\hat m \ap_{-1}^{P}
\ap_{-1}^{P}\ap_{-1}^{P}| n-3,n-3 \rangle+ 2\underbrace{\ap_{-1}^{P}%
\ap_{-2}^{P}| n-3,n-3 \rangle}_{\mathrm{irrelevant } \leftarrow(\ref{d1p1}%
)}+\underbrace{\ap_{-1}^{P}\ap_{-1}^{P}| n-2,n-3 \rangle}_{\mathrm{irrelevant
} \leftarrow(\ref{d2p1})},
\end{align}
and we conclude that both class of states are irrelevant at high energies
\begin{align}
\ap_{-1}^{P}\ap_{-1}^{P}| n-2,n-3 \rangle, \quad\mbox{and} \quad\ap_{-1}%
^{P}\ap_{-1}^{P}\ap_{-1}^{P}| n-3,n-3 \rangle\rightarrow\mathrm{irrelevant } .
\label{n=1}%
\end{align}

\subsubsection{Irrelevance of the states with odd numbers of $\ap_{-1}^{P}$ and
$\ap_{-1}^{L}$}

The previous result (\ref{n=1}) allows us to use mathematical induction. We
need to prove that if both
\begin{align}
\overbrace{\ap_{-1}^{P}\cdots\ap_{-1}^{P}}^{2k-2}| n-2k+2,n-2k+1 \rangle,
\quad\mbox{and} \quad\overbrace{\ap_{-1}^{P}\cdots\ap_{-1}^{P}}^{2k-1}|
n-2k+1,n-2k+1 \rangle\rightarrow\mathrm{irrelevant } , \label{preresult}%
\end{align}
then
\begin{align}
\overbrace{\ap_{-1}^{P}\cdots\ap_{-1}^{P}}^{2k}| n-2k,n-2k-1 \rangle,
\quad\mbox{and} \quad\overbrace{\ap_{-1}^{P}\cdots\ap_{-1}^{P}}^{2k+1}|
n-2k-1,n-2k-1 \rangle \rightarrow\mathrm{irrelevant }.%
\end{align}
The proof is consisted of computing the following two types of spurious
states:
\begin{align}
&  L_{-1}\overbrace{\ap_{-1}^{P}\cdots\ap_{-1}^{P}}^{2k-1}| n-2k,n-2k-1
\rangle\nn\\
&  =\hat m \overbrace{\ap_{-1}^{P}\cdots\ap_{-1}^{P}}^{2k}| n-2k,n-2k-1
\rangle+(2k-1)\underbrace{\overbrace{\ap_{-1}^{P}\cdots\ap_{-1}^{P}}^{2k-2}
\ap_{-2}^{P}| n-2k,n-2k-1 \rangle}_{\mathrm{irrelevant } \leftarrow
(\ref{preresult})}\nn\\
&  +\underbrace{\overbrace{\ap_{-1}^{P}\cdots\ap_{-1}^{P}}^{2k-1} |
n-2k+1,n-2k-1 \rangle}_{\mathrm{irrelevant } \leftarrow(\ref{preresult})} ,
\label{devenp1}%
\end{align}
and
\begin{align}
&  L_{-1}\overbrace{\ap_{-1}^{P}\cdots\ap_{-1}^{P}}^{2k}| n-2k-1,n-2k-1
\rangle\nn\\
&  =\hat m \overbrace{\ap_{-1}^{P}\cdots\ap_{-1}^{P}}^{2k+1}| n-2k-1,n-2k-1
\rangle+ 2k\underbrace{\overbrace{\ap_{-1}^{P}\cdots\ap_{-1}^{P}}%
^{2k-1}\ap_{-2}^{P}| n-2k-1,n-2k-1 \rangle}_{\mathrm{irrelevant }
\leftarrow(\ref{preresult})}\nn\\
&  +\underbrace{\overbrace{\ap_{-1}^{P}\cdots\ap_{-1}^{P}}^{2k}| n-2k,n-2k-1
\rangle}_{\mathrm{irrelevant } \leftarrow(\ref{devenp1})}.
\end{align}
Therefore, both type of states
\begin{align}
\overbrace{\ap_{-1}^{P}\cdots\ap_{-1}^{P}}^{2k}| n-2k,n-2k-1 \rangle,
\quad\mbox{and} \quad\overbrace{\ap_{-1}^{P}\cdots\ap_{-1}^{P}}^{2k+1}|
n-2k-1,n-2k-1 \rangle,
\end{align}
can be ignored.
Using the identity
\begin{equation}
\ap_{-1}^{L}=\ap_{-1}^{P}+\ap_{-1}^{(L-P)},
\end{equation}
we conclude that
\begin{align}
\overbrace{\ap_{-1}^{P}\cdots\ap_{-1}^{P}\ap_{-1}^{L}\cdots\ap_{-1}^{L}}^{2k}|
n-2k,n-2k-1 \rangle, \quad\mbox{and} \quad\overbrace{\ap_{-1}^{P}%
\cdots\ap_{-1}^{P}\ap_{-1}^{L}\cdots\ap_{-1}^{L}}^{2k+1}| n-2k-1,n-2k-1
\rangle,
\end{align}
are also irrelevant, because the naive dimension of $\ap_{-1}^{(L-P)}$ is zero.

\subsection{Linear relations}

In this section we rederive the linear relations among 4-point functions
obtained in \cite{CHLTY}.

So far, we have shown that if a state is not of this form,
\begin{align}
\overbrace{\ap_{-1}^{P}\cdots\ap_{-1}^{P}\ap_{-1}^{L}\cdots\ap_{-1}^{L}}^{2k}|
n-2k,n-2k \rangle, \label{relevant}%
\end{align}
then the state is irrelevant. Furthermore, it indicates that a combination
\begin{align}
&  (\ap_{-1}^{L}\ap_{-1}^{L}-\ap_{-1}^{P}\ap_{-1}^{P})\overbrace{\ap_{-1}%
^{P}\cdots\ap_{-1}^{P}\ap_{-1}^{L}\cdots\ap_{-1}^{L}}^{2k}| n-2k-2,n-2k-2
\rangle\nn\\
&  =(2\ap_{-1}^{P}\ap_{-1}^{(L-P)}+\ap_{-1}^{(L-P)}\ap_{-1}^{(L-P)}%
)\overbrace{\ap_{-1}^{P}\cdots\ap_{-1}^{P}\ap_{-1}^{L}\cdots\ap_{-1}^{L}}%
^{2k}| n-2k-2,n-2k-2 \rangle,
\end{align}
is also irrelevant. By
definition,
$\overbrace{\ap_{-1}^{T}\cdots\ap_{-1}^{T}}^{n}| 0 \rangle$ is relevant. Now
we derive relations among the above type of states and see that all of them
are relevant. Hereafter, we use the notation
\begin{align}
{\mathcal{T}}^{(n,2m,q)},
\end{align}
to represent scattering amplitudes corresponding to the states
\begin{align}
|n, 2m, q\rangle\equiv\left(  \a^{T}_{-1}\right)  ^{n-2m-2q}\left(
\a^{P}_{-1}\right)  ^{2m}\left(  \a_{-2}^{P}\right)  ^{q} |0; k\rangle.
\end{align}

\subsubsection{Relation 1}

Consider the spurious state
\begin{align}
L_{-2}\overbrace{\ap_{-1}^{T}\cdots\ap_{-1}^{T}}^{n-2}| 0 \rangle=\left(
\frac{1}{2}(\underbrace{\ap_{-1}^{T}\ap_{-1}^{T}}_{\mathrm{relevant}%
}+\underbrace{\ap_{-1}^{L}\ap_{-1}^{L}-\ap_{-1}^{P}\ap_{-1}^{P}}%
_{\mathrm{irrelevant } })+\hat m \ap_{-2}^{P}+\underbrace{\ap_{-3}\cdot
\ap_{1}+\cdots}_{\mathrm{irrelevant } }\right)  \overbrace{\ap_{-1}^{T}%
\cdots\ap_{-1}^{T}}^{n-2}| 0 \rangle.
\end{align}
We see that $\overbrace{\ap_{-1}^{T}\cdots\ap_{-1}^{T}}^{n-2}\ap_{-2}^{P}| 0
\rangle$ is relevant and
\begin{align}
{\mathcal{T}}^{(n,0,1)}=-\frac{1}{2\hat m}{\mathcal{T}}^{(n,0,0)}.
\end{align}
Using mathematical induction, we find that $\overbrace{\ap_{-1}^{T}%
\cdots\ap_{-1}^{T}}^{n-2q}\overbrace{\ap_{-2}^{P}\cdots\ap_{-2}^{P}}^{q}| 0
\rangle$ are relevant and
\begin{align}
\label{rel1}{\mathcal{T}}^{(n,0,q)}=\left(  \frac{-1}{2\hat m}\right)
^{q}{\mathcal{T}}^{(n,0,0)}.
\end{align}

\subsubsection{Relation 2}

Consider another class of spurious states
\begin{align}
&  L_{-1}\overbrace{\ap_{-1}^{T}\cdots\ap_{-1}^{T}}^{n-2q-2}\ap_{-1}%
^{P}\overbrace{\ap_{-2}^{P}\cdots\ap_{-2}^{P}}^{q}| 0 \rangle\nn\\
&  =(\hat m \ap_{-1}^{P}+\ap_{-2}\cdot\ap_{1}+\underbrace{\ap_{-3}\cdot
\ap_{2}+\cdots}_{\mathrm{irrelevant } })\overbrace{\ap_{-1}^{T}\cdots
\ap_{-1}^{T}}^{n-2q-2}\ap_{-1}^{P}\overbrace{\ap_{-2}^{P}\cdots\ap_{-2}^{P}%
}^{q}| 0 \rangle\nn\\
&  =\hat m\overbrace{\ap_{-1}^{T}\cdots\ap_{-1}^{T}}^{n-2q-2}\ap_{-1}%
^{P}\ap_{-1}^{P}\overbrace{\ap_{-2}^{P}\cdots\ap_{-2}^{P}}^{q}| 0
\rangle+\underbrace{\overbrace{\ap_{-1}^{T}\cdots\ap_{-1}^{T}}^{n-2q-2}%
\overbrace{\ap_{-2}^{P}\cdots\ap_{-2}^{P}}^{q+1}| 0 \rangle}%
_{\mathrm{relevant}}+\mathrm{irrelevant } .
\end{align}
It shows that $\overbrace{\ap_{-1}^{T}\cdots\ap_{-1}^{T}}^{n-2q-2}\ap_{-1}%
^{P}\ap_{-1}^{P}\overbrace{\ap_{-2}^{P}\cdots\ap_{-2}^{P}}^{q}| 0 \rangle$ is
relevant and
\begin{align}
{\mathcal{T}}^{(n,2,q)}=-\frac{1}{\hat m}{\mathcal{T}}^{(n,0,q+1)}.
\end{align}
Using mathematical induction again, we find that $\overbrace{\ap_{-1}%
^{T}\cdots\ap_{-1}^{T}}^{n-2m-2q}\overbrace{\ap_{-1}^{P}\cdots\ap_{-1}^{P}%
}^{2m}\overbrace{\ap_{-2}^{P}\cdots\ap_{-2}^{P}}^{q}| 0 \rangle$ are relevant
and
\begin{align}
{\mathcal{T}}^{(n,2m,q)}=\left(  \frac{-(2m-1)}{\hat m}\right)  \cdots\left(
\frac{-3}{\hat m}\right)  \left(  \frac{-1}{\hat m}\right)  {\mathcal{T}%
}^{(n,0,q+m)}. \label{rel2}%
\end{align}

\subsection{Final result}

Because we know the irrelevance of other states, the flipping of $P$ to $L$ is
justified for relevant states. Therefore all relevant states take the form
\begin{align}
\overbrace{\ap_{-1}^{T}\cdots\ap_{-1}^{T}}^{n-2m-2q} \overbrace{\ap_{-1}%
^{P}\cdots\ap_{-1}^{P}\ap_{-1}^{L}\cdots\ap_{-1}^{L}}^{2m} \overbrace
{\ap_{-2}^{P}\cdots\ap_{-2}^{P}\ap_{-2}^{L}\cdots\ap_{-2}^{L}}^{q}| 0 \rangle,
\end{align}
and their amplitudes are related to that of the reference state $|n\rangle =\overbrace
{\ap_{-1}^{T}\cdots\ap_{-1}^{T}}^{n}| 0 \rangle$ by
\begin{align}
{\mathcal{T}}^{(n,2m,q)} = \left(  \frac{-(2m-1)}{\hat m}\right)
\cdots\left(  \frac{-3}{\hat m}\right)  \left(  \frac{-1}{\hat m}\right)
\left(  \frac{-1}{2\hat m}\right)  ^{m+q} {\mathcal{T}}^{(n,0,0)}.
\label{main}%
\end{align}
This is finally our main result of the previous paper \cite{CHLTY}.

\section{Choice of gauge}

In the last section, we have proved, under the assumption (\ref{assumption}), that, in the
high energy limit, all 4-point correlation functions are linearly related and
remarkably there is a unique function for every family. It suggests that there
exists such a choice of basis for physical particles, where only one particle
in the same family gives a non-zero scattering amplitude. Though such a basis
could in principle be given in any gauge, we found that there exists a
suitable gauge where one immediately realizes that only one physical particle
survives in the high energy limit for every mass level.

This gauge is naturally spanned by DDF positive norm states \cite{DDF}. DDF
positive norm states are created by acting DDF operators,
\begin{equation}
A_{n}^{i}=\frac{1}{2\pi}\int_{0}^{2\pi}d\tau\dot{X}^{i}(\tau)e^{inX^{+}(\tau
)}d\tau,\text{ }i=1,...,24, \label{DDF operator}%
\end{equation}
on the tachyonic ground state $|0,p_{0}\rangle$, where the tachyonic momentum
is chosen as $p_{0}^{\mu}=(0,\cdots,0,\sqrt{2})$. It is well-known that such
states span the whole spectrum of physical positive norm states,
\begin{equation}
A_{-n_{1}}^{i_{1}}A_{-n_{2}}^{i_{2}}\cdots A_{-n_{m}}^{i_{m}}|0,p_{0}\rangle,
\end{equation}
in a frame where the momentum of these states takes the form of $p^{\mu}%
=p_{0}^{\mu}-Nk_{0}^{\mu}$, with $k_{0}=(-1,0,\cdots,0,1)/\sqrt{2}$ and $N$
representing the level of states. Since this construction of physical positive
norm states naturally picks up a gauge, we simply call this gauge the DDF
gauge. This gauge is characterized by the condition,
\begin{equation}
k_{0}\cdot\alpha_{n}|\mathrm{physical\,\,state\rangle=0,}%
\end{equation}
for $n>0$. This condition guarantees that, in this gauge, physical states
contain only oscillators whose polarizations are perpendicular to $k_{0}$.
(Note that $k_{0}$ is perpendicular to itself because it is a null vector.)

After short calculation, we can rewrite the DDF states in terms of the usual
Fock space representation in any Lorentz flame. For example, for the 1st
massive particles ($m^{2}=2$), we have,
\begin{align}
A_{-1}^{i}A_{-1}^{j}|0,p_{0}\rangle &  \rightarrow\left(  \ap_{-1}^{i}\ap_{-1}%
^{j}+\delta^{ij}\left(  -\frac{1}{2\sqrt{2}}\ap_{-2}^{(L-P)}+\frac{1}%
{4}\ap_{-1}^{(L-P)}\ap_{-1}^{(L-P)}\right)  \right)  |0,p\rangle
,\label{DDF11}\\
A_{-2}^{i}|0,p_{0}\rangle &  \rightarrow\left(  \ap_{-2}^{i}-\sqrt{2}\ap_{-1}%
^{i}\ap_{-1}^{(L-P)}\right)  |0,p\rangle, \label{DDF12}%
\end{align}
and for the 2nd massive particles ($m^{2}=4$),
\begin{align}
&  A_{-1}^{i}A_{-1}^{j}A_{-1}^{k}|0,p_{0}\rangle\rightarrow\nn\\
&  \left(  \ap_{-1}^{i}\ap_{-1}^{j}\ap_{-1}^{k}+\left(  \delta^{ij}%
\ap_{-1}^{k}+\delta^{ki}\ap_{-1}^{j}+\delta^{jk}\ap_{-1}^{i}\right)  \left(
-\frac{1}{4}\ap_{-2}^{(L-P)}+\frac{1}{8}\ap_{-1}^{(L-P)}\ap_{-1}%
^{(L-P)}\right)  \right)  |0,p\rangle,\label{DDF21}\\
&  A_{-2}^{i}A_{-1}^{j}|0,p_{0}\rangle\rightarrow\nn\\
&  \left(  \ap_{-2}^{i}\ap_{-1}^{j}-\ap_{-1}^{i}\ap_{-1}^{j}\ap_{-1}%
^{(L-P)}+\delta^{ij}\left(  -\frac{1}{3}\ap_{-3}^{(L-P)}+\frac{1}{2}%
\ap_{-2}^{(L-P)}\ap_{-1}^{(L-P)}-\frac{1}{6}\ap_{-1}^{(L-P)}\ap_{-1}%
^{(L-P)}\ap_{-1}^{(L-P)}\right)  \right)  |0,p\rangle,\nn\\
& \label{DDF22}\\
&  A_{-3}^{i}|0,p_{0}\rangle\rightarrow\left(  \ap_{-3}^{i}-\frac{3}{2}\ap_{-2}%
^{i}\ap_{-1}^{(L-P)}+\ap_{-1}^{i}\left(  -\frac{3}{4}\ap_{-2}^{(L-P)}-\frac
{9}{8}\ap_{-1}^{(L-P)}\ap_{-1}^{(L-P)}\right)  \right)  |0,p\rangle.
\label{DDF23}%
\end{align}
The polarization $e^{(L-P)}$ of oscillator $\ap_{-n}^{(L-P)}$ emerges as a
covariantized version of the null vector $k_{0}$. This explicit expressions of
the physical positive norm states and our previous counting rules of naive
dimension (\ref{naivedim}) imply that only the particles
(\ref{DDF11}) and (\ref{DDF21}) with $i,j,k=T$ are
relevant in the family at mass levels $m^{2}=2$ and $m^{2}=4$, respectively.
Actually, this kind of structure is generic in this gauge. Noticing that the
DDF operator (\ref{DDF operator}) is at most linear in oscillators
$\ap_{-n}^{i}$, and the naive dimension of $\ap_{-n}^{(L-P)}$ is $0$, we
should assign naive dimension $1$ to $A_{-n}^{T}$ and $0$ to others.
Therefore, we can conclude that the unique state,
\begin{equation}
A_{-1}^{T}\cdots A_{-1}^{T}|0,p_{0}\rangle, \label{DDFresult}%
\end{equation}
constructed by only one operator, $A_{-1}^{T}$, survives in the high energy
limit, under the same assumption as we have made in the previous section.

The existence of the DDF gauge where only the transverse and the
$(L-P)$-polarization are needed is a direct implication of the fact that DDF
states include all inequivalent physical states (up to Lorentz
transformations). In this gauge we can easily see that the ratio of high
energy amplitude is simply $1:0:\cdots:0$, while in other choices of gauge
(and thus different choices of basis), they might be different. For instance,
in the gauge where we use only the transverse and longitudinal polarizations,
the ratio is 8:1:-1:-1 for $m^{2} = 4$ (for the basis chosen in
\cite{ChanLee1}).

\section{Validity of the assumption}

\label{Check of the assumption}

In the previous sections, we have studied behaviors of stringy amplitudes in
the high-energy limit, based on the assumption made in sec. \ref{Spurious}. In
this section, we shall check the validity of the assumption (\ref{assumption}%
). At first sight, one might think that the validity of the assumption is
rather trivial. If there is no suppression of the energy orders of the
high-energy amplitudes with the particle $| n \rangle$, we can regard their
naive dimensions as the true ones, hence the assumption is valid. But we know
that, in some situations, the high-energy amplitudes are suppressed and their
true energy orders become less than the naive ones. Typically, such a
suppression occurs when the other vertices contain $\ap^{P}_{-1}$ or
$\ap^{L}_{-1}$. Originally, Wick-contracted terms between tensor parts of
different vertex operators give subleading contributions compared to those
from Wick-contracted terms between tensors and exponents. However, if the
(naive) leading contributions turn out to cancel and the amplitudes are
suppressed, it is very likely that the true leading contributions to a
multi-tensor amplitude will be given by these tensor-tensor contraction terms.
In that situation, the naive energy orders of amplitudes by dimensional
analysis, which is based on Wick-contractions between tensors and exponents,
might fail. Therefore, we need to check the validity of our assumption in
order to confirm the main results (\ref{relevant}), (\ref{main}) and
(\ref{DDFresult}).

{}To be specific, we shall explicitly check the assumption in a representative
example, where the four-point function consists of two spin-two tensors (at
mass level $m^{2}=2$) and two tachyons. In particular, we shall fix $V_{1}$ as
a physical state which consists $\ap_{-1}^{L}$, and we can expect that the
amplitude gets suppressed. We shall calculate the following amplitude
\begin{equation}
\mathcal{T}=\int\prod_{i=1}^{4}dx_{i}\langle V_{1}V_{2}V_{3}V_{4}%
\rangle,\nonumber
\end{equation}
where
\[
V_{1}\equiv\partial X^{T_{1}}\partial X^{L_{1}}e^{ik_{1}X},
\]
and $V_{3},V_{4}$ are tachyon vertices. For $V_{2}$, we consider the following
three cases:
\begin{align}
\text{Case1}  &  \text{: \ \ }V_{2}=\partial X^{T_{2}}\partial X^{T_{2}%
}e^{ik_{2}X},\label{Case1}\\
\text{Case2}  &  \text{: \ \ }V_{2}=\partial X^{T_{2}}\partial X^{(L_{2}%
-P_{2})}e^{ik_{2}X},\label{Case2}\\
\text{Case3}  &  \text{: \ \ }V_{2}=\partial X^{(L_{2}-P_{2})}\partial
X^{(L_{2}-P_{2})}e^{ik_{2}X}. \label{Case3}%
\end{align}
Notice that by our energy-counting rule (\ref{naivedim}), the four-point
amplitudes associated with the three cases should have naive dimensions five,
four and three, respectively. While it is conceivable that the true leading
energy order for the first case, due to the presence of $\partial X^{L_{1}}$,
should be no greater than three; it is not trivial to see whether the true
leading orders of the second and the third cases are really less than that of
the first case such that our assumption (\ref{assumption}) can be justified.
For this reason, we need to perform a sample calculation, based on the
saddle-point method\footnote{For details of the saddle-point method, see sec.
5 of \cite{CHLTY}.}.

The amplitude with two spin-two tensors and two tachyons is given by\footnote{
We have employed the standard $SL(2,R)$ gauge fixing, $x_{1}=0$,
$x_{3}=1$, $x_{4}=\infty$, such that $k_{14}$ ,$k_{24}$ and $k_{34}$ do not
appear in this discussion.}
\[
\int_{-\infty}^{\infty}dxu(x)e^{-Kf(x)},
\]
where
\begin{align}
&  K=-k_{1}\cdot k_{2},\qquad\tau=-\frac{k_{2}\cdot k_{3}}{k_{1}\cdot k_{2}%
},\\
&  f(x)=\ln x-\tau\ln(1-x),
\end{align}
and the function $u(x)$ consists of three contributions with different energy
orders. It is convenient to make the following decomposition,
\begin{equation}
u(x)\equiv u_{I}(x)+u_{II}(x)+u_{III}(x),
\end{equation}
and
\begin{align}
u_{I}(x)\equiv &  (e^{T_{1}}\cdot k_{23})(e^{L_{1}}\cdot k_{23})(e^{A}\cdot
k_{13})(e^{B}\cdot k_{13}),\label{uone}\\
u_{II}(x)\equiv &  -\frac{1}{x^{2}}\Big[(e^{T_{1}}\cdot e^{A})(e^{L_{1}}\cdot
k_{23})(e^{B}\cdot k_{13})+(e^{T_{1}}\cdot e^{B})(e^{L_{1}}\cdot k_{23}%
)(e^{A}\cdot k_{13})\nn\\
&  +(e^{L_{1}}\cdot e^{A})(e^{T_{1}}\cdot k_{23})(e^{B}\cdot k_{13}%
)+(e^{L_{1}}\cdot e^{B})(e^{T_{1}}\cdot k_{23})(e^{A}\cdot k_{13})\Big],\\
u_{III}(x)\equiv &  \frac{1}{x^{4}}\Big[(e^{T_{1}}\cdot e^{A})(e^{L_{1}}\cdot
e^{B})+(e^{T_{1}}\cdot e^{B})(e^{L_{1}}\cdot e^{A})\Big],
\end{align}
where $e_{\mu}^{A}e_{\nu}^{B}$ is the polarization tensor of the second
particle and
\begin{align}
k_{23}  &  \equiv\frac{k_{2}}{x}+k_{3},\\
k_{13}  &  \equiv\frac{k_{1}}{x}-\frac{k_{3}}{1-x},
\end{align}
are linear combinations of the momenta, which come from the Wick-contraction
with $e^{ik\cdot X}$s. The function $u_{I}(x)$ is the part of $u(x)$ which does
not contain terms from tensor-tensor contraction. $u_{II}(x)$ and $u_{III}$
come from the terms with one tensor-tensor contraction and with two
tensor-tensor contractions, respectively. Note that, when we use the
saddle-point approximation and substitute its saddle-point value
$x_{0}=1/(1-\tau)$ for moduli parameter $x$, the inner products $e^{L_{1}%
}\cdot k_{23}$ and $e^{L_{2}}\cdot k_{13}$ get suppressed and their true
energy orders become one. In Case 1, in the zeroth-order contribution of the
saddle-point approximation, we can see the expected suppression for both
$u_{I}(x_{0})$ and $u_{II}(x_{0})$, and $u_{III}(x_{0})$ is identically zero
because the polarizations are orthogonal to $e^{L_{1}}$. Then, we conclude
that the true energy order of the amplitude is really three, after checking
that there is no further suppression between the zeroth-order contribution of
$u_{I}(x)$ and the first-order contribution of $u_{I}(x)$ in the saddle-point
approximation. In view of this, for the validity of the assumption, we need to
make sure that the true energy orders of Case 2 and 3 are really less than
three. Indeed, due to the existence of $e^{L_{1}}$ in $V_{1}$, $u_{I}(x)$ for
both Case 2 and 3 also get suppressed and have energy order two and one,
respectively. In general, $u_{I}(x)$ always shares a common pattern of
suppression for different choices of $V_{2}$, because it is of a factorized
form. Furthermore, for Case 2 and 3, naive energy orders (hence, true energy
orders) of $u_{II}(x)$ and $u_{III}(x)$ are no greater than
two\footnote{Actually, they do not get suppressed in this example.}. Thus, in
this example, our assumption is valid.

Now we can apply our results to this example and check one of their
consequences. For example, if we consider the vertex
\begin{align}
\mbox{Case 4: } \quad V_{2}=\partial X^{T_{2}}\partial X^{L_{2}} e^{i k_{2}
X},
\end{align}
according to our results, this amplitude should have true energy order less
than that of Case 1, namely, three. At first sight, it does not seem to happen
because the true energy order of $u_{\II}(x_{0})$ is actually four. However,
explicit calculation shows that the first-order contribution of $u_{I}(x)$ in
the saddle-point approximation (miraculously) cancels the leading contribution
of $u_{\II}(x_{0})$. Then, the true energy order of Case 4 is two as we have
predicted. The following table summarizes the energy orders of the $u$'s for
all four cases. The right arrow stands for suppressions or cancellations.
\[%
\begin{array}
[c]{|c|c|c|c|}\hline
& u_{I}(x) & u_{\II}(x) & u_{\III}(x)\\\hline
\mathrm{Case 1} & E^{5}\rightarrow E^{3} & E^{3}\rightarrow E^{1} & 0\\\hline
\mathrm{Case 2} & E^{4}\rightarrow E^{2} & E^{2} & E^{0}\\\hline
\mathrm{Case 3} & E^{3}\rightarrow E^{1} & E^{1} & E^{-1}\\\hline
\mathrm{Case 4} & E^{6}\rightarrow E^{4} & E^{4} & E^{2}\\\cline{2-3}
& \multicolumn{2}{c|}{E^{4}+E^{4}\rightarrow E^{2}} & \\\hline
\end{array}
\]

We can generalize the above argument to any other multi-tensor amplitudes as
far as no other particle contains polarizations $e^{T_{i}}$. In order to make
discussion simple, we choose the physical states for $V_{1}$, $V_{3}$ and
$V_{4}$ in the DDF gauge. In this gauge, the polarizations of positive-norm states
will consist of $e^{T}$ and $e^{(L-P)}$ only (e.g. Eqs.(\ref{DDF11}%
)-(\ref{DDF23})). Consequently, suppressions which we have seen in the
previous example do not happen and true leading amplitudes are simply given by
the zeroth-order saddle-point approximation. In particular, we expect that
$u_{I}(x_{0})$, analogously defined as in Eq.(\ref{uone}), should give the
leading contribution to the high-energy amplitudes. Because $u_{I}(x_{0})$
depends on the polarization of $V_{2}$ only through factors of $e^{A}\cdot
k_{13}$, its true energy order is exactly same as what we expect from the
naive dimension of $V_{2}$. The only exception might happen if we consider
amplitudes with subleading particles. For subleading particles in the DDF gauge,
there are several terms which make the leading contributions to $u_{I}(x_{0}%
)$, for example, $\ap_{-2}^{T}| 0,p \rangle$ and $\sqrt{2}\ap_{-1}^{T}%
\ap_{-1}^{(L-P)}| 0,p \rangle$ in Eq.(\ref{DDF12}). If there is an unexpected
cancellation among these terms, we cannot rely on the above argument and must
judge the assumption on a case-by-case basis. However, cancellation itself is
easily checked by calculating one factor of $u_{I}(x_{0})$, which is related
to the polarization of the subleading particle. Therefore, as we did in the
previous example, after checking that there is no such a cancellation, we can
conclude that the assumption is valid. Thus, we can expect that our assumption
is widely valid as far as the other particles do not contain polarizations of
$e^{T_{i}}$.

On the other hand, if we consider amplitudes with $e^{T_{i}}$, our assumption
easily breaks down. For example, if an amplitude contains one photon with the
polarization $e^{T_{i}}$, the amplitude vanishes unless it has another
particle with polarization $e^{T_{i}}$. Therefore, the leading particle should
be $\del X^{T_{i}}(\del X^{T})^{n-1}e^{ik\cdot X}$, not $(\del X^{T}%
)^{n}e^{ik\cdot X}$. Thus, our assumption (\ref{assumption}) is not valid.
However, if we replace the state $| n \rangle$ by $| n-1 \rangle^{\prime
}\equiv\ap_{-1}^{T_{i}}(\ap_{-1}^{T})^{n-1}| 0 \rangle$ and repeat the same
argument in section \ref{Spurious}, one should be able to derive similar
results based on this reference state.

\section*{Acknowledgment}

The authors thank Masako Asano and Hiroyuki Hata for helpful discussions. 
This work is supported in part by
the National Science Council, Taiwan, R.O.C and National Center for Theoretical
Sciences, Hsinchu, Taiwan, R.O.C (grant NSC 94-2119-M-002-001).

\end{document}